\documentclass[final,3p,times,11pt]{elsarticle}

\usepackage[utf8]{inputenc}
\usepackage{amsmath,amssymb}
\usepackage{amsfonts}
\usepackage{enumerate}
\usepackage{bm}
\usepackage{graphicx}

\makeatletter
\def\ps@pprintTitle{%
 \let\@oddhead\@empty  
 \let\@evenhead\@empty
 \def\@oddfoot{\centerline{\thepage}}%
 \let\@evenfoot\@oddfoot}
\makeatother

\newcommand{\p}{\partial}

\newcommand{\ex}{\bm{\hat{e}}_1}
\newcommand{\ey}{\bm{\hat{e}}_2}
\newcommand{\ez}{\bm{\hat{e}}_3}

\newcommand{\magn}{\bm{m}}

\newcommand{\DM}{D}

\newcommand{\Anisotropy}{K}

\newcommand{\dmscaled}{\epsilon}

\newcommand{\ldw}{\ell_{\rm w}}
\newcommand{\ldk}{\ell_{S}}
\newcommand{\hf}{\textstyle{\frac{1}{2}}}

\newcommand{\dx}{\mathrm{d}x}
\newcommand{\dy}{\mathrm{d}y}

\begin{document}

\begin{frontmatter}

\title{The profile of chiral skyrmions of small radius}
\author{Stavros Komineas}
\address{Department of Mathematics and Applied Mathematics, University of Crete, 70013 Heraklion, Crete, Greece}
\author{Christof Melcher}
\address{Department of Mathematics \& JARA Fundamentals of Future Information Technology, RWTH Aachen University, 52056 Aachen, Germany}
\author{Stephanos Venakides}
\address{Department of Mathematics, Duke University, Durham, NC, USA}
\date{\today}

\begin{abstract}
Chiral skyrmions are stable particle-like solutions of the Landau-Lifshitz equation for ferromagnets with the Dzyaloshinskii-Moriya (DM) interaction, characterized by a topological number.
We study the profile of an axially symmetric skyrmion and give exact formulas for the solution of the corresponding far-field and near-field equations, in the asymptotic limit of small DM parameter (alternatively large anisotropy).
The matching of these two fields leads to a formula for the skyrmion radius as a function of the DM parameter.
The derived  solutions show the different length scales which are present in the skyrmion profiles.
The picture is thus created of a chiral skyrmion that is born out of a Belavin-Polyakov solution with an infinitesimally small radius, as the DM parameter is increased from zero.
The skyrmion retains the Belavin-Polyakov profile over and well-beyond the core before it assumes an exponential decay; the profile of an axially-symmetric Belavin-Polyakov solution of unit degree plays the role of the universal core profile of chiral skyrmions.
\end{abstract}


\end{frontmatter}

%%%%%%%%%%%%%%%%%%%%%%%%%%%%%%
\section{Introduction}

Magnetic skyrmions were predicted to be stabilized in ferromagnets with the Dzyaloshinskii-Moriya (DM) interaction \cite{BogdanovYablonskii_JETP1989,BogdanovHubert_JMMM1994}.
They have been observed in Dzyaloshinskii-Moriya (DM) materials and techniques have been developed for individual skyrmions to be created and annihilated in a controlled manner \cite{RommingHanneken_Science2013}.
Skyrmions are examples of topological magnetic solitons in ferromagnetic films that exhibit particle-like behavior in the sense that they are localized robust entities both regarding their statics and their dynamical behavior.
This makes them attractive for theoretical studies in order to understand details of their behavior while it also gives them a strong potential for applications \cite{FertReyrenCros_NRM2017}. 

Magnetic solitons \cite{KosevichIvanovKovalev_PR1990}, such as magnetic bubbles and vortices \cite{MalozemoffSlonczewski,HubertSchaefer}, have been investigated theoretically and experimentally, and their global features (such as topology and qualitative morphology) have been observed to an extent.
It is though only in recent years that experimental techniques have been developed that offer sufficient resolution for the observation of detailed features of the skyrmion profile \cite{RommingKubetzka_PRL2015,McGrouther_NJP2016,LeonovWiesendanger_NJP2016, BoulleVogel_nnano2016,KovacsBorkovski_PRL2017,ShibataTokura_PRL2017}.
The details of the skyrmion profile determine to a large extent and sometimes crucially the properties of the skyrmion \cite{BuettnerLemeshBeach_srep2018} and is thus essential for the manipulation of individual skyrmions.

Skyrmions can be found as solutions of the Landau-Lifshitz equation in the presence of DM interaction by numerical methods.
The existence of such solutions has been rigorously proved \cite{Melcher_PRSA2014,Li_Melcher_JFA2018}, but so far no analytic formula for the skyrmion profile has grown out of rigorous mathematical reasoning.
Instead, an ad-hoc ansatz based on explicit domain wall profiles \cite{Braun_PRB1994} has been suggested and is widely used to examine structural and dynamic properties, see, e.g., \cite{RommingKubetzka_PRL2015, Zhou_NCOMM_2015,BuettnerLemeshBeach_srep2018}.
Further trial profiles have been tested to this end \cite{TejoRiverosChubykalo_srep2018,TomaselloGuslienkoFinocchio_PRB2018}.

In this paper we derive formulas for the skyrmion profile by employing asymptotic methods that give analytic approximations for the solutions of the Landau-Lifshitz equation.
Our methods are valid for the case of small DM parameter or large anisotropy 
and they can readily be extended to the case of a large external field.
The derived solutions show the detailed features and the different length scales which are present in the skyrmion profile..
The role of the DM interaction for the existence of skyrmion solutions and the role of the Belavin-Polyakov solution as a universal limit of skyrmion profiles are revealed.
The approach of this paper fails for large skyrmion radius.
The latter case can be handled using different techniques \cite{KomineasMelcherVenakides_arXiv2019b}.

The availability of mathematically derived formulas will facilitate the comparison of experimentally observed profiles, particularly focusing on some of their special features, and may be useful for a variety of other purposes \cite{MantelMuratovSimon_PRB2019,MantelMuratovSimon_arXiv2019}.
Specifically, the skyrmion profile enters in an essential way in formulas for dynamical phenomena \cite{PapanicolaouTomaras_NPB1991,KomineasPapanicolaou_PhysD1996}, for example, skyrmion translation and rotation modes \cite{SchuetteGarst_PRB2014}, and it is crucial for quantitative calculations.
Finally, the methods developed in this paper are potentially  useful in the search for solutions of the Landau-Lifshitz equation in cases of skyrmion dynamics.

The paper is arranged as follows.
The basic equations are presented in section 2.
The far field is analyzed in section 3, where it is given as a series.
Section 4 is devoted to the calculation of the near field.
The matching of the near and far fields is performed in section 5 and the skyrmion radius is calculated as a function of the small DM parameter in section 6.
A summary and discussion of results are presented in section 7.
Finally, an outline for an existence proof of the skyrmion is presented in Appendix A.

%%%%%%%%%%%%%%%%%%%%%%%%%%%%%%
\section{The basic equations}

\subsection{Magnetization vector}

We assume a ferromagnetic material as a two-dimensional system lying on the $xy$-plane.
The micromagnetic structure is described via the magnetization vector $\magn=\magn(x,y)$ with a fixed magnitude normalized to unity, $\bm{m}^2=1$.
We will assume a ferromagnet with exchange interaction, a Dzyaloshinskii-Moriya (DM) interaction, and an anisotropy of the easy-axis type perpendicular to the film, governed by the normalized energy
\begin{equation}
E_\dmscaled(\magn)=\int \left[ \hf \p_\mu\magn\cdot\p_\mu\magn + \hf (1-m_3^2) + \dmscaled\, \bm{\hat{e}}_\mu\cdot (\p_\mu\magn\times\magn) \right]\,\dx
\end{equation}
where summation over repeated indices $\mu=1,2$ is implied and $\ex,\ey,\ez$ are the unit vectors for the magnetization in the respective directions.
Static magnetization fields are local minimizers of $E_{\dmscaled}$ satisfying the normalized Landau-Lifshitz equation 
\begin{equation} \label{eq:LL0}
\magn \times \bm{h}_{\dmscaled}=0
\end{equation}
where the effective field 
\begin{equation}
    \bm{h}_{\dmscaled}= \p_\mu\p_\mu\magn + m_3 \ez - 2\dmscaled\, \bm{\hat{e}}_\mu\times\p_\mu\magn
\end{equation}
is minus the variational gradient of $E_\dmscaled =E_\dmscaled(\magn)$.
We measure lengths in units of the domain wall width $\ldw = \sqrt{A/K}$, where $A$ is the exchange and $\Anisotropy$ the anisotropy parameter.
The equation contains a single parameter
\begin{equation}  \label{eq:parameter}
\dmscaled = \frac{\ldk}{\ldw} = \frac{\DM}{2\sqrt{A\Anisotropy}}
\end{equation}
defined via an additional length scale of this model $\ldk = \DM/(2\Anisotropy)$, where $\DM$ is the DM parameter (in Ref.~\cite{BogdanovHubert_PSS1994}, a parameter which differs from $\epsilon$ only by a constant factor has been introduced).
The lowest energy (ground) state is the spiral for $\dmscaled > 2/\pi$ and the ferromagnetic state for $\dmscaled < 2/\pi$ \cite{BogdanovHubert_JMMM1994,ChovanPapanicolaou_PRB2002}. Isolated chiral skyrmions occur in the ferromagnetic regime as local energy minimizers in a nontrivial homotopy class \cite{BogdanovYablonskii_JETP1989, BogdanovHubert_JMMM1994, BogdanovHubert_JMMM1999, Melcher_PRSA2014, Li_Melcher_JFA2018}.  

Let us consider the angles $(\Theta, \Phi)$ for the spherical parametrization of the magnetization vector, and the polar coordinates $(r,\phi)$ for the film plane.
We assume an axially symmetric skyrmion with $\Phi = \phi+\pi/2$ and $\Theta=\Theta(r)$, called a Bloch skyrmion.
All subsequent calculations remain valid (actually identical) if, instead of the bulk DM term in Eq.~\eqref{eq:LL0} and a Bloch skyrmion, we consider one of the types of skyrmions obtained for different types of the DM interaction (corresponding to different crystallographic classes) discussed in Ref.~\cite{BogdanovYablonskii_JETP1989} and reproduced also graphically in Ref.~\cite{LeonovWiesendanger_NJP2016}.
For example, these include a N\'eel skyrmion with $\Phi=\phi$ for an interfacial DM term.
The equation for the profile $\Theta=\Theta(\rho)$ 
    \begin{equation}\label{eq:thetaODE}
\Theta''+\frac{\Theta'}{r}-\frac{\sin(2 \Theta)}{2 r^2}    -  \frac{\sin(2 \Theta)}{2} 
+ 2\dmscaled \frac{ \sin^2 \Theta}{r} = 0
\end{equation}
with boundary conditions $\Theta(0)=\pi$ and $\lim_{r \to \infty}\Theta(r)=0$ is the same for all typed of skyrmions, therefore the following calculations apply equally to all of them.

%%%%%%%%%%%%%%%%%%%%%%%%%%%%%%
\subsection{Stereographic projection}

We define the field
\begin{equation}
    u(r) = \tan\frac{\Theta(r)}{2}
\end{equation}
which is the modulus of the stereographic projection of the magnetization vector.
The equation for $u$ reads
\begin{equation}  \label{eq:u}
u'' + \frac{u'}{r} + \frac{u}{r^2} \frac{u^2-1}{u^2+1} - \frac{2 u u'^2}{u^2+1} + \frac{u^2-1}{u^2+1} u + 4\dmscaled \frac{u^2}{r (u^2+1)} = 0.
\end{equation}
where the prime denotes differentiation with respect to $r$.
A well-known solution of this equation has been obtained for the case where only the exchange interaction is present, that is, when  the last two terms in Eq.~\eqref{eq:u} are absent.
This is the axially-symmetric Belavin-Polyakov (BP) solution of unit degree
\begin{equation}  \label{eq:BPskyrmion}
u(r) = \frac{a}{r},
\end{equation}
where $a$ is an arbitrary complex constant and $|a|$ gives the radius of the skyrmion core \cite{BelavinPolyakov_JETP1975}.

In order to remove the singularity at the origin, we define the field
\begin{equation}  \label{eq:v}
    v(s) = s u,\qquad s=\frac{r}{2}
\end{equation} 
which transforms Eq. \eqref{eq:u} to the quasilinear equation (linear in the highest derivative; $\dot v=\frac{dv}{ds}$)
\begin{equation}  \label{eq:vs_aniso}
\ddot{v} + \frac{\dot{v}}{s}\, \frac{3v^2-s^2}{v^2+s^2} - \frac{2 v \dot{v}^2}{v^2 + s^2} + 4\frac{v^2 - s^2}{v^2+s^2} v = - 8\dmscaled \frac{v^2}{v^2+s^2}.
\end{equation}

\begin{figure}[t]
    \centering
    \includegraphics[width=6cm]{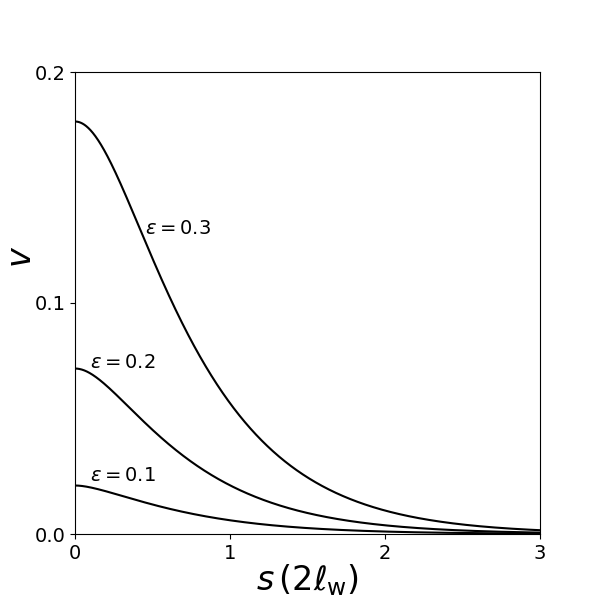}
    \caption{Skyrmion profiles $v(s)$ for three values of the parameter $\dmscaled$, as indicated in each curve.
    The function $v(s)$ is monotonically decreasing and the value at the skyrmion center $v_0=v(0)$ decreases with decreasing $\dmscaled$.
    The coordinate $s$ is measured in units of twice the domain wall width ($2\ldw$).}
    \label{fig:profiles}
\end{figure}

\begin{figure}[t]
    \centering
    \includegraphics[width=4.5cm]{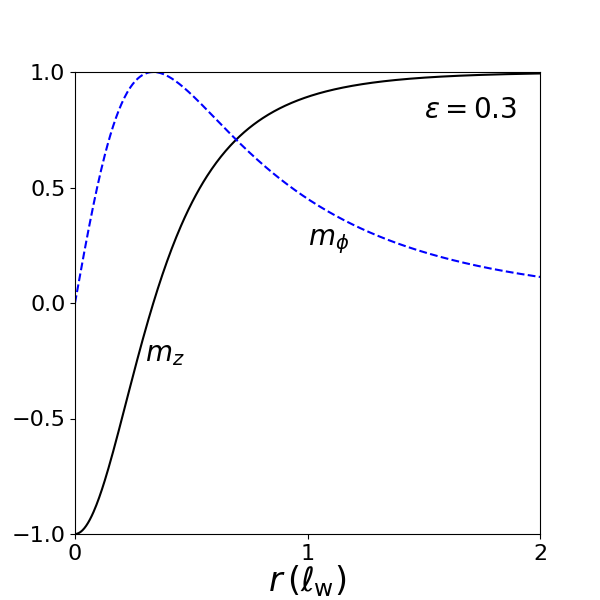}
    \includegraphics[width=4.5cm]{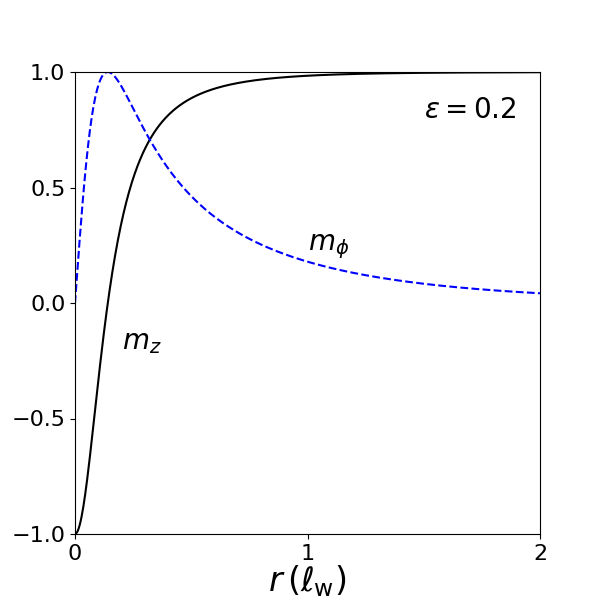}
    \includegraphics[width=4.5cm]{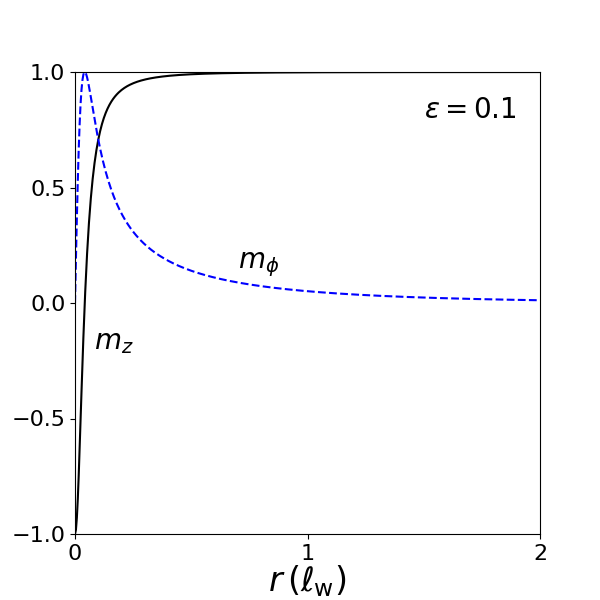}
    \caption{The magnetization components $(m_\phi,m_z)$ for the skyrmion profiles plotted in Fig.~\ref{fig:profiles}.
    The skyrmion radius is decreasing with $\dmscaled$.
    The coordinate $r$ is measured in domain wall width ($\ldw$) units.}
    \label{fig:magn}
\end{figure}

In order to display the findings of our analysis, we find numerically the skyrmion profiles for various values of the parameter $\dmscaled$ by employing a relaxation algorithm for solving the Landau-Lifshitz equation \eqref{eq:LL0} \cite{KomineasPapanicolaou_PRB2015a} as well as by solving numerically Eq.~\eqref{eq:vs_aniso} using a shooting method.
In the latter method we start integration at $s=0$ choosing $v(s=0)=v_0$, where $v_0$ is chosen arbitrarily, $\dot{v}(s=0)=0$ and we integrate up to large values of $s$.
We seek and actually find a $v_0$ such that $v(s)$ decreases monotonically tending to zero for large $s$.
The results of both methods agree, but we only obtain a result by the shooting method for small $\dmscaled$, while the relaxation method converges extremely slowly and is not precise for $\dmscaled \lesssim 0.2$.
Fig.~\ref{fig:profiles} shows the skyrmion profiles $v(s)$ for three values of $\dmscaled$ and leads to the following three observations.
(1) Field $v(s)$ takes a finite non-zero value at $s=0$ indicating that the stereographic field $u(r)$ diverges as $1/r$ at $r\to 0$, similar to the BP skyrmion of Eq.~\eqref{eq:BPskyrmion}.
(2) The value $v(0)$ decreases as $\dmscaled$ decreases, e.g., for larger anisotropy or smaller DM parameter.
(3) The profiles $v(s)$ are monotonically decreasing indicating a faster decay of $u(r)$ in comparison to the BP solution \eqref{eq:BPskyrmion}.

Fig.~\ref{fig:magn} depicts the magnetization vector components for the skyrmion profiles of Fig.~\ref{fig:profiles}.
The radius of the skyrmion is clearly seen to decrease with decreasing $\dmscaled$ in agreement with our calculations below.

Motivated by the observations on Fig.~\ref{fig:profiles}, we seek a small-$\epsilon$ asymptotic solution of Eq.~\eqref{eq:vs_aniso} for which $v(s)$ is small and tends to zero at spatial infinity.
A further assumption, which is found to be consistent with our results, is that we can neglect the term containing $\dot{v}^2$ as compared to the DM term on the right of the equation.
We thus impose the asymptotic assumptions 
\begin{equation}\label{eq:asympt_conditions}
 v\ll 1,   \quad \dot v\ll\sqrt{v\dmscaled},  \;\; \mbox{for} \,    \dmscaled\ll 1.
\end{equation}
 Then Eq.~\eqref{eq:vs_aniso} simplifies to the semilinear equation (linear in all derivatives),
\begin{equation}  \label{eq:vs_aniso_semilinear}
\ddot{v} + \frac{\dot{v}}{s}\, \frac{3v^2-s^2}{v^2+s^2}  + 4\frac{v^2 - s^2}{v^2+s^2} v = - 8\dmscaled \frac{v^2}{v^2+s^2}.
\end{equation}
We solve Eq.~\eqref{eq:vs_aniso_semilinear} asymptotically, by obtaining two separate linear equations, one for the far field and one for the near field, and matching their solutions through an overlap subdomain of the independent variable $s$.

The matching condition yields the relation between the boundary value $v(0)=v_0$ and the small parameter $\dmscaled$.
The significance of this relationship  is that {\it the value $v_0$ equals the radius of the skyrmion in the asymptotic limit}. 
Indeed, $\Theta=\frac{\pi}{2}$ at the skyrmion radius, thus $v=s$.
In the limit $\dmscaled\to 0$, the value of $v$ at which $v(s)=s$, converges to $v_0$ (see Eq.~\eqref{eq:nearField_match} ).

We verify a posteriori, that the profile obtained satisfies the conditions \eqref{eq:asympt_conditions}.

\section{The far field}

The assumption $u\ll 1$ defines the range of the far-field. Eq.~\eqref{eq:u} simplifies to the modified Bessel equation 
\begin{equation}  \label{eq:u_farfield}
r^2u'' + ru' - (r^2+1)u = 0.
\end{equation}
The decaying solution of this equation is proportional to the modified Bessel function of the second kind $K_1(r)$ \cite{AbramowitzStegun}.
A rearrangement of this gives
\begin{equation}  \label{eq:u_decay}
u = \frac{1}{s} + (2\gamma-1) s + 2 \sum_{n=1}^\infty \frac{(\ln s - \ln n) s^{2n-1}}{(n-1)!\, n!} - 2 \sum_{n=2}^\infty \frac{\xi_n s^{2n-1}}{(n-1)!\, n!},\qquad s=\frac{r}{2}
\end{equation}
where $\gamma \approx 0.5772$ is the Euler-Mascheroni constant and
\begin{equation}
\xi_n = \frac{1}{2} (\psi(n)+\psi(n+1))-\ln n, \ \ \  \ \ \ \
\psi(n)=-\gamma+\sum_{k=1}^{n-1}\frac{1}{k}, \ \ \ \ \qquad n=2,3,4,\ldots. 
\end{equation}
The spatial variable $s=\frac{r}{2}$, used here and below instead of $r$ makes our equations and formulae simpler. 
The asymptotic behavior of the solution as $s\to\infty$ is given by 
\begin{equation}  \label{eq:farField_asymptotic-u}
    u\sim\sqrt{\frac{\pi}{s}}\,e^{-2s}\left(1+\frac{3}{4s}+\cdots  \right).
\end{equation}

For the variable $v(s)$, the differential equation corresponding to Eq.~\eqref{eq:u_farfield} reads
\begin{equation}  \label{eq:vs_farfield}
\ddot{v} - \frac{\dot{v}}{s} - 4 v = 0,
\end{equation}
with exact solution given by the series
\begin{equation}  \label{eq:v4_decay}
    v = 1 + (2\gamma-1) s^2 + 2 \sum_{n=1}^\infty \frac{(\ln s - \ln n) s^{2n}}{(n-1)!\, n!} - 2 \sum_{n=2}^\infty \frac{\xi_n s^{2n}}{(n-1)!\, n!}
\end{equation}
and asymptotic behavior as $s\to\infty$ given by
\begin{equation}  \label{eq:farField_asymptotic}
    v\sim\sqrt{\pi s}\, e^{-2s}\left(1+\frac{3}{4s}+\cdots  \right).
\end{equation}

\section{The near field}

In the near field, we replace $v$ in Eq.~\eqref{eq:vs_aniso_semilinear} 
  with its initial value $v_0$.
The approximation is valid for the range of $s$ over which $v_0-v\ll v_0$. 
The equation obtained in this way,
\begin{equation}  \label{eq:vs_nearField}
\ddot{v} + \frac{\dot{v}}{s}\, \frac{3v_0^2-s^2}{v_0^2+s^2} = -4\frac{v_0^2 - s^2}{v_0^2+s^2} v_0 -8\dmscaled \frac{v_0^2}{v_0^2+s^2},
\end{equation}
is integrable by the integrating factor
\[
\mu(s) = \frac{s^3}{(v_0^2+s^2)^2}.
\]
We obtain
\[
\frac{d}{ds} \left[ \frac{s^3}{(v_0^2+s^2)^2} \dot{v} \right] = -4\frac{v_0^2 - s^2}{(v_0^2+s^2)^3} v_0 s^3 -8\dmscaled \frac{v_0^2 s^3}{(v_0^2+s^2)^3},
\]
which integrates to
\[
\frac{s^4}{(v_0^2+s^2)^2}\frac{\dot{v}}{s} = -2\dmscaled \frac{s^4}{(v_0^2+s^2)^2} 
+ 2v_0 \left[ \ln\left( 1 + \frac{s^2}{v_0^2} \right) - \frac{v_0^2+2s^2}{(v_0^2+s^2)^2} s^2 \right].
\]
We finally have
\begin{equation}  \label{eq:nearField_vdot}
\frac{\dot{v}}{s} = -2\dmscaled + 2v_0 \left[ \frac{(v_0^2+s^2)^2}{s^4} \ln\left( 1 + \frac{s^2}{v_0^2} \right) - \frac{v_0^2+2s^2}{s^2} \right].
\end{equation}
The constant of integration has been judiciously chosen to eliminate the fourth-order singularity $s^{-4}$ on the right.

We proceed to the integration of Eq.~\eqref{eq:nearField_vdot}.
We make the change of variables $w(\tau)=v/v_0$ and $\tau=s^2/v_0^2$.
Then Eq.~\eqref{eq:nearField_vdot} becomes
\begin{equation}  \label{eq:nearField_wdot}
\frac{dw}{d\tau} = -\dmscaled v_0 + v_0^2 \left[ \ln(1+\tau) + 2 \frac{\ln(1+\tau)}{\tau} + \frac{\ln(1+\tau)}{\tau^2} - \frac{1}{\tau} - 2 \right].
\end{equation}
We integrate Eq.~\eqref{eq:nearField_wdot} and obtain
\begin{equation}  \label{eq:nearField_wtau}
w(\tau) = 1 - \dmscaled v_0 \tau + v_0^2 \left[ \tau \ln(1+\tau) - \frac{1}{\tau} \ln(1+\tau) + 1 - 2{\rm Li}_2(-\tau)  - 3\tau \right]
\end{equation}
where ${\rm Li}_n(x)$ is the polylogarithm function and the constant of integration has been chosen so that $w(0)=1$.
Keeping the dominant terms for large $\tau$ we have
\begin{equation}
w(\tau) = 1 - \left(\dmscaled + 3v_0 \right) v_0\tau + v_0^2 \tau \ln(\tau)
\end{equation}
which, in the original variables, gives the near field
\begin{equation}  \label{eq:nearField_match}
v_N(s) = v_0 - \left( \dmscaled + 3v_0 + 3 v_0 \ln v_0 \right) s^2 + 2v_0 s^2 \ln s.
\end{equation}

\begin{figure}[t]
    \centering
    \includegraphics[width=4.5cm]{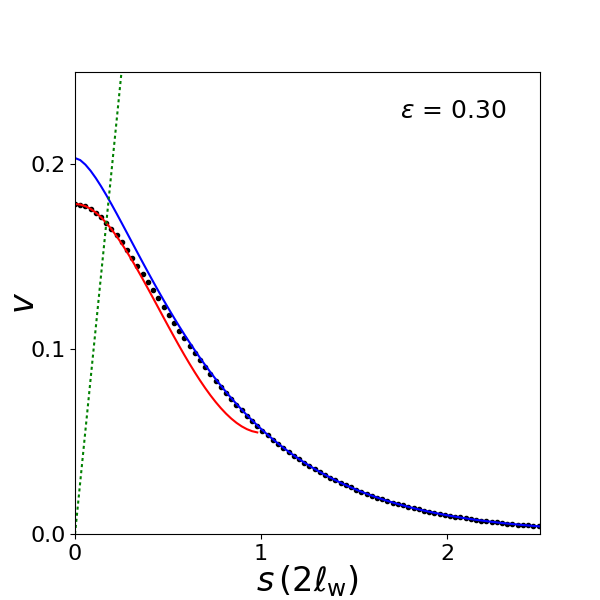}
    \includegraphics[width=4.5cm]{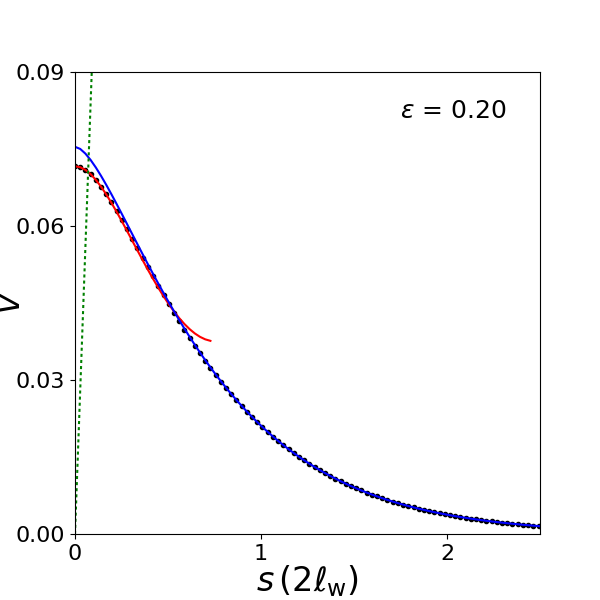}
    \includegraphics[width=4.5cm]{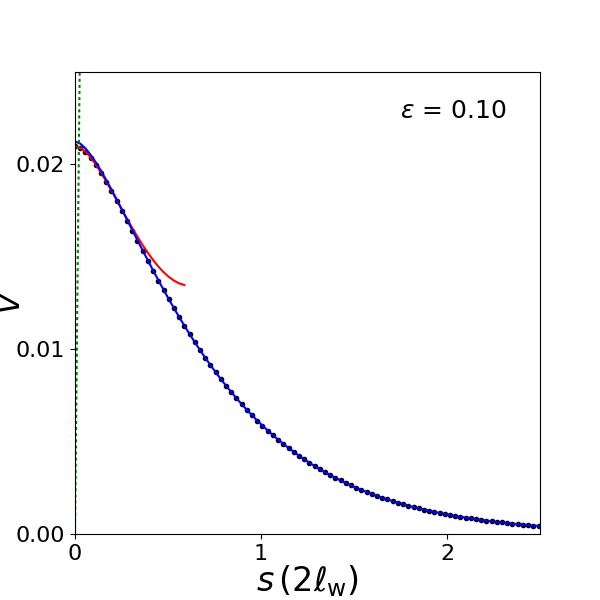}
    \caption{The numerically calculated profiles of skyrmions for three values of the parameter $\dmscaled$, as indicated in each figure, are shown by black dots.
    The blue solid line shows the far-field approximation of the skyrmion profile given by Eq.~\eqref{eq:v4_decay}, which is a solution of Eq.~\eqref{eq:vs_farfield}.
    The arbitrary factor of the far field has been chosen such that it fits the numerical data at large $s$.
    The red solid line shows the near-field approximation of Eq.~\eqref{eq:nearField_vdot}, or Eq.~\eqref{eq:nearField_wtau}, which is a solution of Eq.~\eqref{eq:vs_nearField}.
    The free parameter $v_0$ of the near field has been chosen so that it agrees with the numerical data at $s=0$.
    The green dotted line is $v=s$ and it cuts the skyrmion profile at the skyrmion radius.
    The skyrmion radius is decreasing for decreasing $\dmscaled$ and it is seen graphically that it is approximately equal to $v_0$ for small $\dmscaled$.}
    \label{fig:profiles_matching}
\end{figure}

Fig.~\ref{fig:profiles_matching} shows the numerically calculated skyrmion profiles for three values of $\dmscaled$ together with the corresponding approximations for the far field in Eq.~\eqref{eq:v4_decay} and the near field in Eq.~\eqref{eq:nearField_wtau}.
The arbitrary constant of the near and far fields are chosen to agree with the numerical solutions.
The near-field approximation turns to an increasing function of $s$ past its domain of validity and this part is not plotted in the figure.

When $\dmscaled$ is small enough a matching between the near field and the far field can be done through the near-field Eq.~\eqref{eq:nearField_match} and the far-field Eq.~\eqref{eq:farField_match}, shown below, without the use of numerical data.

%%%%%%%%%%%%%%%%%%%%%%%%%%%%%%
\section{Asymptotic matching}
\label{sec:matching}

The far-field equation has been derived under the assumptions $v_0^2 \ll s^2$ and  $\dot v\ll s$.
A condition for the validity of the near-field approximation, consistent with inequality \eqref{eq:asympt_conditions} is $\dmscaled s^2 \ll v_0$.
We also require $v_0<\dmscaled$, which is automatically satisfied in our calculation below.
Thus, both the far-field and the near-field asymptotic solutions are valid in the internal layer
\begin{equation}
v_0^2 \ll s^2 \ll \frac{v_0}{\dmscaled} < 1,\qquad \dmscaled\to 0.
\end{equation}

We observe that both the far-field expression obtained from Eq.~\eqref{eq:v4_decay}, when terms of order $O(s^4 \ln s)$ are neglected,
\begin{equation}  \label{eq:farField_match}
 v_F(s) = a_0 \left[ 1+(2\gamma-1) s^2 + 2 s^2 \ln s \right]
\end{equation}
and the near-field Eq.~\eqref{eq:nearField_match}
are linear combinations of a constant term with $s^2\ln s$ and $s^2$.
Matching the coefficients of these terms in the two equations yields
\begin{equation}\label{eq:epsilon_v0_relation1}
\dmscaled = -2 v_0 \left(\gamma + 1 + \ln v_0 \right),
\end{equation}
or, more compactly
\begin{equation}  \label{eq:asymptoticMatching}
\dmscaled = -2 v_0 \ln\left(\frac{v_0}{\alpha}\right),\qquad \alpha = e^{-(\gamma + 1)} \approx 0.2065.
\end{equation}
The last equation determines implicitly the value of $v_0$ as a function of the parameter $\dmscaled$.

\begin{figure}[t]
\centering
\includegraphics[width=6cm]{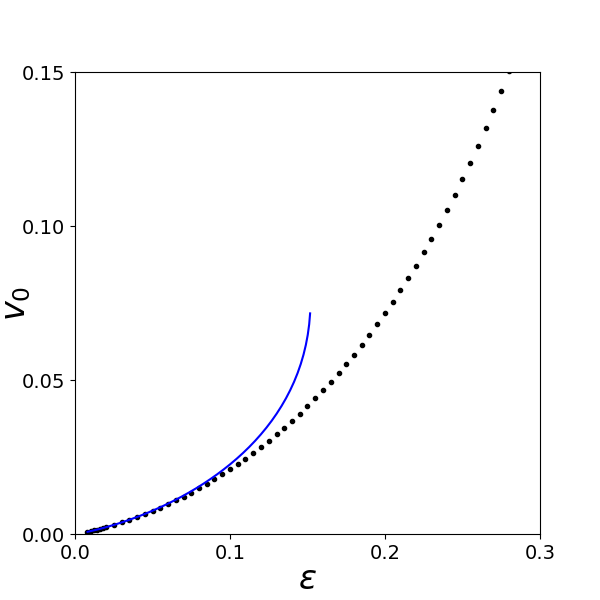}
\caption{Numerically calculated values for $v_0=v(0)$ for various values of $\dmscaled$ are shown by black dots.
The approximation \eqref{eq:asymptoticMatching} obtained by the asymptotic matching is shown by the blue solid line.
The approximation formula does not have solutions for $\dmscaled \gtrsim 0.15$. 
This graph may also be viewed as an approximation for the skyrmion radius $R$ because $R \approx 2v_0$ for $v_0 \ll 1$.}
\label{fig:v0-a}
\end{figure}

We have solved numerically by a shooting method Eq.~\eqref{eq:vs_aniso} for various values of $\dmscaled$ and found the values $v_0=v(0)$ for the skyrmion profiles.
In Fig.~\ref{fig:v0-a} we plot the value $v_0$ as a function of $\dmscaled$ and compare the numerical data with formula \eqref{eq:asymptoticMatching}.
The agreement is excellent for small values of $\dmscaled \lesssim 0.1$.
For these values of $\dmscaled$ the skyrmion profile is that of the Belavin-Polyakov soliton \eqref{eq:BPskyrmion} for a range of $s$ far exceeding the skyrmion radius, while the profile tail is exponentially decaying for large $s$ as shown in Eq.~\eqref{eq:farField_asymptotic}.

The equations for both the near and far field in which terms of order $O(s^4)$ or $O(s^4 \ln s)$ are neglected and their derivatives are
\begin{equation}  \label{eq:skyrmionProfile}
 v(s) = v_0 \left[ 1+ 2 s^2 \ln s+(2\gamma-1) s^2 +\cdots \right],
 \qquad  \dot{v}(s)=4v_0[s\ln s+\gamma+\cdots]
\end{equation}
each holding in its respective domain, $v_0\ll s$ for the far field and $s^2\ln s\ll1$ for the near field.

%%%%%%%%%%
\section{Skyrmion radius}

\subsection{Skyrmion radius from asymptotic matching}
A formula for the skyrmion radius for small values of $\dmscaled$ can be determined from the above results.
In order to see this note that the skyrmion radius is the solution of the equation $v(s)=s$.
From Eq.~\eqref{eq:skyrmionProfile} we find that $v(s) \approx v_0$, for small values of $s$, that is, the skyrmion radius $R$ is approximately at $s=v_0$.
Since lengths for $s$ are measured in units of $2\ldw$ we have 
\begin{equation}\label{eq:R=2v_0}
    R=2 v_0
\end{equation}
 in units $\ldw$. 
We can now write Eq.~\eqref{eq:asymptoticMatching} for the skyrmion radius as
\begin{equation}  \label{eq:skyrmionRadius}
\dmscaled = -R \ln\left(\frac{R}{2\alpha}\right),\qquad \dmscaled \ll 1.
\end{equation} 
For $R\ll 1$ the logarithm in Eq.~\eqref{eq:skyrmionRadius} may be approximated by the Lambert $W$ function \cite{Hoorfar_Hassani}, for which the equation can be inverted in terms of elementary functions, i.e.,
\begin{equation} \label{eq:approx_radius}
R \approx -  \frac{\dmscaled}{\ln \left( \frac{\dmscaled}{2 \alpha} \right)} \approx -  \frac{\dmscaled}{\ln \dmscaled}
\end{equation}
for $\dmscaled \ll 1$.
For values of $\dmscaled \gtrsim 0.12$ Eq.~\eqref{eq:skyrmionRadius} overestimates the skyrmion radius while it gives no result for $\dmscaled \gtrsim 0.15$.
Numerical results show that the skyrmion radius increases with increasing $\dmscaled$ and it diverges to infinity for $\dmscaled \to 2/\pi$, in agreement with theoretical results.
The latter behavior in not captured by formula \eqref{eq:skyrmionRadius}. 
The study of the regime for $\dmscaled \gtrsim 0.15$ using methods similar to those introduced in this paper is left to another study.

Restoring physical constants in Eq.~\eqref{eq:approx_radius} we have
\begin{equation}  \label{eq:radiusDK}
R \approx -\frac{\frac{\DM}{2\Anisotropy}}{\ln\left(\frac{\DM}{2\sqrt{A\Anisotropy}}\right)},\qquad \DM \ll \sqrt{A\Anisotropy}.
\end{equation}
This shows that the skyrmion radius decreases for decreasing DM parameter or increasing anisotropy.
In the limit $\DM\to 0$ or $\Anisotropy \to \infty$, the skyrmion radius $R$ goes to zero.
A form of Eq.~\eqref{eq:skyrmionRadius} which is more instructive as it implicates the length scales of this model, is
\begin{equation}
    R \approx -\frac{\ldk}{\ln\left(\frac{\ldk}{\ldw} \right)}.
\end{equation}

We will finally apply the above formulas and give specific examples.
The regime of small $\dmscaled$ can be obtained for small enough values of the DM parameter $\DM$.
Furthermore, for a regime where the skyrmion has larger radius we assume a small anisotropy $\Anisotropy$, which gives a large $\ldw$ (domain wall width scale).
As a first example let us choose $\dmscaled=0.3$ so that the formulas for the near and the far field give good approximations as seen in Fig.~\ref{fig:profiles_matching}.
From Fig.~\ref{fig:v0-a} we have a skyrmion radius $R \approx 2 v_0 \approx 0.36\ldw$.
As a second example let us choose $\dmscaled=0.1$ which falls within the range of validity of the asymptotic matching and Eq.~\eqref{eq:asymptoticMatching}.
We then have a skyrmion radius $R \approx 2 v_0 \approx 0.04\ldw$.

\subsection{Universal core profile and energy asymptotics}

Our approach indicates that the profile of an axially-symmetric Belavin-Polyakov 
solution of unit degree plays the role of the universal core profile of chiral skyrmions $\magn_\dmscaled$ in the asymptotic regime $\dmscaled \ll 1$. The formula for the skyrmion radius in Eq.~\eqref{eq:skyrmionRadius} identifies a scaling law for the size of the skyrmion core.  It is therefore natural to rescale space by the approximate skyrmion radius $R_\dmscaled = - \frac{\dmscaled}{\ln \dmscaled}$ from Eq. \eqref{eq:approx_radius} to obtain a non-collapsing family of magnetization fields 
\begin{equation} \label{eq:log_scaling}
    \hat{\magn}_\dmscaled(y) = \magn_\dmscaled\left( x \right)
    \quad \text{where} \quad y = x/R_\dmscaled.
\end{equation}
Taking into account Eq.~\eqref{eq:v}, rescaling $t=s/R_{\dmscaled}$ entails the following rescaling of $v=v_\dmscaled$ 
\begin{equation}
    \hat{v}_\dmscaled(t)= \frac{v_\dmscaled(R_\dmscaled t)}{R_\dmscaled} \quad \text{where} \quad t=\frac{|y|}{2}.
\end{equation}
According to Eqs. \eqref{eq:skyrmionProfile} and \eqref{eq:R=2v_0} we have 
$\hat{v}_\dmscaled(t) \to \frac{1}{2}$ uniformly for bounded $t$. Consequently, the fields $\hat{\magn}_\dmscaled$ are uniformly approximated on compact subsets of $\mathbb{R}^2$ by the
normalized Belavin-Polyakov solution $\bm{\phi}$, i.e., the rotated stereographic map used in \cite{Melcher_PRSA2014, Doering_Melcher_2017}.
Upgrading this approximation property in terms of integral norms supports the conjectured asymptotics of minimal skyrmion energies $E_\dmscaled=E_\dmscaled(\magn_\dmscaled)$ 
\begin{equation} \label{eq:energy_asymptotics}
E_\dmscaled - 4 \pi \sim \frac{\dmscaled^2}{\ln \dmscaled}
\end{equation}
in the regime $\dmscaled \ll 1$. 
The upper bound (with leading constant $c=1$ in the present scaling) has been established in \cite{Doering_Melcher_2017} by means of appropriate trial fields. 
A matching lower bound requires an ansatz-free argument that is not at hand, but we shall outline a heuristics based on scaling and convergence.
Eq.~\eqref{eq:log_scaling} yields the rescaled energy
\begin{equation}\label{eq:res_energy}
\hat{E}_\dmscaled(\hat \magn) = \int_{\mathbb{R}^2} \left[ \frac{1}{2} \p_\mu\hat \magn \cdot \p_\mu\hat \magn + \frac{\dmscaled^2}{\ln \dmscaled} \left(\frac{1-\hat{m}_3^2}{2 \ln \dmscaled}  - \bm{\hat{e}}_\mu\cdot (\p_\mu\hat{\magn}\times\hat{\magn}) \right) \right]\, \dy
\end{equation}
with $\hat{E}_\dmscaled(\hat{\magn}_\dmscaled) = E_\dmscaled(\magn_\dmscaled)$.
The key property of the integrand of $\hat{E}_\dmscaled(\hat{\magn}_\dmscaled)$ is that the prefactor to anisotropy and DM term is proportional to the expected energy gain in Eq.~\eqref{eq:energy_asymptotics}.
The Derrick-Pohozaev identity \cite{Derrick} 
implies a balance of anisotropy and DM term
\begin{equation} \label{eq:equipart}
 \int_{\mathbb{R}^2}   \frac{1-\hat{m}_{3,\dmscaled}^2}{\ln \dmscaled}\, \dy=
 \int_{\mathbb{R}^2}  \bm{\hat{e}}_\mu\cdot (\p_\mu \hat{\magn}_\dmscaled \times \hat{\magn}_\dmscaled) 
 \, \dy.
\end{equation}
Claiming strong convergence $\hat{\magn}_\dmscaled - \bm{\phi} \to 0$ in $\dot{H}^1 \cap L^4(\mathbb{R}^2)$, which is consistent with the limited decay properties of the Belavin-Polyakov solution $\bm{\phi}$, the arguments in \cite{Doering_Melcher_2017} ensure convergence of the renormalized DM term, i.e., 
\begin{equation}
 \lim_{\dmscaled \searrow 0}   \int_{\mathbb{R}^2}  \bm{\hat{e}}_\mu\cdot \left( \p_\mu \hat{\magn}_\dmscaled \times \hat{\magn}_\dmscaled \right)
 \, \dy= \int_{\mathbb{R}^2}  \bm{\hat{e}}_\mu\cdot \left[ \p_\mu \bm{\phi} \times \left(\bm{\phi}-\bm{\hat{e}}_3\right) \right]
 \, \dy= -8\pi. 
\end{equation}
The resulting bounds on the anisotropy term in Eq.~\eqref{eq:equipart} feature in particular the well-known logarithmic divergence of mass for degree one solitons in the pure exchange model. But more importantly, going back to Eq.~\eqref{eq:res_energy}, we obtain a matching lower energy bound and hence the precise energy asymptotics
\begin{equation}
E_{\dmscaled}= 4\pi \left(1 + \frac{\dmscaled^2}{\ln \dmscaled} \right) + o\left(\frac{\dmscaled^2}{\ln \dmscaled} \right)
\end{equation}
for $\dmscaled \ll 1$, provided of course that minimal energies are attained by axially-symmetric skyrmions. A fully rigorous argument quantifying all approximation steps and providing suitable compactness properties of rescaled skyrmion configurations is beyond the scope of this discussion and deferred to future studies.

%%%%%%%%%%%%%%%%%%%%%%%%%%%%%%
\section{Summary of results and discussion}

We have derived exact formulae by means of asymptotic methods for the profile of a small radius chiral skyrmion in a model with exchange, easy-axis anisotropy and Dzyaloshinskii-Moriya interaction (bulk or interfacial).
In particular, Eq.~\eqref{eq:radiusDK} gives the skyrmion radius in terms of the system parameters, for small skyrmion radius.
Our results create the picture that the chiral skyrmion is born out of a Belavin-Polyakov (BP) solution with infinitesimally small radius as the DM parameter is increased from zero (or the anisotropy parameter is decreased from infinity).
In the asymptotic limit as $\dmscaled$ approaches zero we find that the skyrmion core radius goes to zero.
The BP profile is a good approximation at the skyrmion core and beyond.
The length scale at which the skyrmion profile breaks away from BP tends to infinity if the skyrmion radius is used as unit of length.
To be sure, as $r\to\infty$ the chiral skyrmion profile decays exponentially (unlike the BP skyrmion).
The basis of all calculations is Eq.~\eqref{eq:vs_aniso} for a scalar field $v$ which is related to the polar angle $\Theta$ of the axially-symmetric magnetization field $\magn$ by 
\begin{equation}
    \Theta(r)=2 \arctan\left(\frac{v(s)}{s} \right) \quad \text{where} \quad s=\frac{r}{2}.
\end{equation}
The field $v$ is derived from the modulus of the stereographic projection of the magnetization vector, transformed in a way that the BP solution of unit degree given by $2 \arctan(a/r)$ with a free scaling factor $a$ is expressed as a constant function $v \equiv a/2$. 

The profile $v(s)$ at large distances (skyrmion tail), which we call the far field,  has a form defined solely by exchange and anisotropy and is therefore exponentially approaching the perpendicular magnetization direction (ferromagnetic state).
The length scale of this part of the profile is the domain wall width $\ldw$.
The profile in this region is given by the series in Eq.~\eqref{eq:v4_decay} with decaying asymptotic form in Eq.~\eqref{eq:farField_asymptotic}.

The profile in the central region (skyrmion core) has the form of a BP solution.
When the parameter $\dmscaled$, given in Eq.~\eqref{eq:parameter}, is small (i.e., for small DM parameter or for large anisotropy) we find a modification to the BP profile as a solution of Eq.~\eqref{eq:vs_nearField}.
We thus obtain a precise form of the profile in the region of the skyrmion core and beyond that as $\dmscaled\to 0$.
We call this profile, given in Eq.~\eqref{eq:nearField_wtau} and approximated in Eq.~\eqref{eq:nearField_match}, the near field.

By matching the formulas for the skyrmion profile obtained at the central region (near field) and at large distances (far field) we obtain the value of a parameter directly connected to the skyrmion radius as a function of $\dmscaled$.
The result, given in Eq.~\eqref{eq:skyrmionRadius}, shows that the skyrmion radius goes to zero as $\dmscaled\to 0$.
The dependence of the radius on $\dmscaled$ supports an asymptotic relation for the minimal skyrmion energy, given in Eq.~\eqref{eq:energy_asymptotics}. 
The skyrmion profile approaches the BP profile in the same limit.

The matching of the near and far fields, in the asymptotic limit of $\dmscaled\to 0$, leads to the same equation for both fields, shown in Eqs.~\eqref{eq:nearField_match} and \eqref{eq:farField_match}, not only on the overlap region, but also beyond it.
The asymptotic coincidence  of the two fields reaches back to $s=0$, in agreement to the profile in the rightmost entry of Fig.~\ref{fig:profiles_matching}, in which the far field alone is a good approximation both far and near. 
The contribution of the near-field equation is yet still crucial; it determines the arbitrary factor of the far-field equation, a factor that depends entirely on the strength of the Dzyaloshinskii-Moriya term.
The role of DM interaction for the existence of skyrmion solutions is, thus, revealed. 
It perturbs the Belavin-Polyakov skyrmion near the position $r=0$ in such a way that the matching with the tail of the skyrmion profile is possible.

A subtle technical point pertains to the term that contains the factor $\dot{v}^2$ in Eq.~\eqref{eq:vs_aniso}.
This makes the equation nonlinear in the derivatives of $v$ setting the stage for the dominance  of this term and the unlimited  growth of $v$.
We have used the ``smallness'' conditions \eqref{eq:asympt_conditions} to control this growth; and since the smallness of $\dot{v}$ is enhanced by squaring it, the term becomes negligible. 
In the Appendix, we outline a proof of existence of the skyrmion, under the condition of small $v_0$, in spite of the presence of the potentially growing term in question.

The functional forms that have been derived in this paper can be used as high precision trial functions in computational approaches including further effects such as stray fields \cite{MantelMuratovSimon_PRB2019,MantelMuratovSimon_arXiv2019}.
They may also motivate more focused experimental schemes for measuring the skyrmion profile.

%%%%%%%%%%
\section*{Acknowledgement}
We are grateful to Stefan Bl\"ugel for pointing out the physical significance of chiral skyrmion profiles.
CM gratefully acknowledges financial support by the DFG under the grant no. ME 2273/3-1, and SK a Mercator fellowship as part of the previous grant.
SV gratefully acknowledges financial support by the NSF through contract DMS-1211638.
SK acknowledges funding for this project from the Hellenic Foundation for Research and Innovation (HFRI) and the General Secretariat for Research and Technology (GSRT), under grant agreement No 871.

\appendix 
\section{Existence of the skyrmion} 
We consider the solutions of Eq.~\eqref{eq:vs_aniso} for some fixed value of $\epsilon$, parametrized by the initial condition  $v(0)=v_0>0$. 
The constraint $\dot v(0)=0$ is forced by the lone $s$ in the denominator in Eq.~\eqref{eq:vs_aniso}.
We observe that $\dot{v}/s\to \ddot{v}(0)$ as $s\to 0$, thus, the equation provides the relation 
\begin{equation}
    \ddot{v}(0)= -(v_0+2\dmscaled)<0,
\end{equation} 
forcing the solution $v=v(s,\,v_0)$ to be initially decreasing for all values of $v_0$.
The decreasing $v(s)$ may not have a critical inflection point.
Indeed, multiplying Eq. \eqref{eq:vs_aniso} by $v^2+s^2$,  taking the derivative with respect to $s$  and letting $\dot v=\ddot v=0$ shows the third derivative of $v(s)$ to be positive, contradicting the existence of such an inflection point.  

We consider three sets of initial data $v_0$, corresponding to three possible scenarios for the solution of  Eq.~\eqref{eq:vs_aniso}.
\begin{enumerate}
    \item[$A_1$:] The solution $v(s)$, originally decreasing, reaches a local minimum at a  positive value of $v$, from where it turns upwards.
    The minimum is stable to sufficiently small perturbations of the initial value $v_0$, due to the continuous dependence of the solution on the initial data.
    Consequently, the set $A_1$ of the $v_0$  values for these  solutions is open.   
    \item[$A_2$:] The solution decays hitting the $s-$axis at $s=s_0$, necessarily, transversely [$\dot v(s_0)=0$, combined with $v(s_0)=0$ would violate the uniqueness of solution of Eq. \eqref{eq:vs_aniso} at $(s_0,0)$].
    This transversality and the absence of critical inflection points make this type of solutions  stable to small perturbations of $v_0$; again, this is due to the continuous dependence of the solution on the initial data.   Therefore the set $A_2$ is open.
    \item[$A_3$:] The solution $v(s)$ decays monotonically to zero, as $s\to\infty$.
    By definition, this is a skyrmion solution. As indicated by the solution of the far-field equation, zero is the only possible limit at infinity for a decaying solution.
\end{enumerate}
The sets $A_1, A_2, A_3$ are clearly mutually disjoint and their union is the semiaxis $(0<v_0<\infty)$. 

Solving Eq. \eqref{eq:vs_aniso} numerically for $0 < \epsilon < 2/\pi$, with initial data $v(0)=v_0$ and $\dot v(0)=0$, we find that taking $v_0=v^+$ large enough, gives an initially decreasing solution that exhibits a minimum  ($v^+\in A_1$); taking $v_0=v^-$ small enough, gives a decaying solution that intersects the $s$-axis ($v_-\in A_2$).
If $I=[v^-, v^+]$, then the intersections  $A_1\cap I$ and $A_2\cap I$ are nonempty, open in $I$, disjoint,  proper subsets of $I$; their union is open in $I$, thus, it is a proper subset of $I$.
Necessarily, there exists a value of $v_0$ in  $I$, which belongs to the set $A_3$.
The above argument guarantees that there exists a skyrmion solution of Eq.~\eqref{eq:vs_aniso} for some $v_0\in (v^-,v^+)$.

\bigskip

\bibliographystyle{unsrt}

%\section*{\refname}
\bibliography{references}

\begin{thebibliography}{10}

\bibitem{BogdanovYablonskii_JETP1989}
A.~N. Bogdanov and D.~A. Yablonskii.
\newblock Thermodynamically stable ``vortices'' in magnetically ordered
  crystals. the mixed state of magnets.
\newblock {\em Sov. Phys. JETP}, 68:101--103, 1989.

\bibitem{BogdanovHubert_JMMM1994}
A.~N. Bogdanov and A.~Hubert.
\newblock Thermodynamically stable magnetic vortex states in magnetic crystals.
\newblock {\em J. Magn. Magn. Mater.}, 138:255, 1994.

\bibitem{RommingHanneken_Science2013}
N.~Romming, C.~Hanneken, M.~Menzel, J.~E. Bickel, B.~Wolter, K.~von Bergmann,
  A.~Kubetzka, and R.~Wiesendanger.
\newblock Writing and deleting single magnetic skyrmions.
\newblock {\em Science}, 341(6146):636--639, 2013.

\bibitem{FertReyrenCros_NRM2017}
A.~Fert, N.~Reyren, and V.~Cros.
\newblock Magnetic skyrmions: advances in physics and potential applications.
\newblock {\em Nature Reviews Materials}, 2:17031, 2017.

\bibitem{KosevichIvanovKovalev_PR1990}
A~M Kosevich, B~A Ivanov, and A~S Kovalev.
\newblock Magnetic solitons.
\newblock {\em Physics Reports}, 194(3-4):117--238, 1990.

\bibitem{MalozemoffSlonczewski}
A.~P. Malozemoff and J.~C. Slonczewski.
\newblock {\em Magnetic Domain Walls in Bubble Materials}.
\newblock Academic Press, New York, 1979.

\bibitem{HubertSchaefer}
A.~Hubert and R.~Sch\"afer.
\newblock {\em Magnetic domains}.
\newblock Springer, Berlin, 1998.

\bibitem{RommingKubetzka_PRL2015}
N.~Romming, A.~Kubetzka, C.~Hanneken, K.~von Bergmann, and R.~Wiesendanger.
\newblock Field-dependent size and shape of single magnetic skyrmions.
\newblock {\em Phys. Rev. Lett.}, 114:177203, May 2015.

\bibitem{McGrouther_NJP2016}
D~McGrouther, R~J Lamb, M~Krajnak, S~McFadzean, S~McVitie, R~L Stamps, A~O
  Leonov, A~N Bogdanov, and Y~Togawa.
\newblock Internal structure of hexagonal skyrmion lattices in cubic
  helimagnets.
\newblock {\em New Journal of Physics}, 18(9):095004, sep 2016.

\bibitem{LeonovWiesendanger_NJP2016}
A~O Leonov, T~L Monchesky, N~Romming, A~Kubetzka, A~N Bogdanov, and
  R~Wiesendanger.
\newblock The properties of isolated chiral skyrmions in thin magnetic films.
\newblock {\em New Journal of Physics}, 18(6):065003, may 2016.

\bibitem{BoulleVogel_nnano2016}
Olivier Boulle, Jan Vogel, Hongxin Yang, Stefania Pizzini, Dayane {de Souza
  Chaves}, Andrea Locatelli, Tevfik~Onur Menteş, Alessandro Sala, Liliana~D.
  Buda-Prejbeanu, Olivier Klein, Mohamed Belmeguenai, Yves Roussign{\'{e}},
  Andrey Stashkevich, Salim~Mourad Ch{\'{e}}rif, Lucia Aballe, Michael
  Foerster, Mairbek Chshiev, St{\'{e}}phane Auffret, Ioan~Mihai Miron, and
  Gilles Gaudin.
\newblock {Room-temperature chiral magnetic skyrmions in ultrathin magnetic
  nanostructures}.
\newblock {\em Nat. Nano.}, 11(5):449--454, may 2016.

\bibitem{KovacsBorkovski_PRL2017}
A.~Kov{\'a}cs, J.~Caron, A.~S. Savchenko, N.~S. Kiselev, K.~Shibata, Zi-An Li,
  N.~Kanazawa, Y.~Tokura, S.~Bl{\"u}gel, and R.~E. Dunin-Borkowski.
\newblock Mapping the magnetization fine structure of a lattice of bloch-type
  skyrmions in an fege thin film.
\newblock {\em Applied Physics Letters}, 111(19):192410, 2017.

\bibitem{ShibataTokura_PRL2017}
K.~Shibata, A.~Kov{\'a}cs, N.~S. Kiselev, N.~Kanazawa, R.~E. Dunin-Borkowski,
  and Y.~Tokura.
\newblock Temperature and magnetic field dependence of the internal and lattice
  structures of skyrmions by off-axis electron holography.
\newblock {\em Physical Review Letters}, 118(8):087202, 2017.

\bibitem{BuettnerLemeshBeach_srep2018}
F.~B\"uttner, I~Lemesh, and S.~D.~G. Beach.
\newblock Theory of isolated magnetic skyrmions: From fundamentals to room
  temperature applications.
\newblock {\em Sci. Rep.}, 8:4464, 2018.

\bibitem{Melcher_PRSA2014}
C.~Melcher.
\newblock Chiral skyrmions in the plane.
\newblock {\em Proc. R. Soc. A}, 470:20140394, October 2014.

\bibitem{Li_Melcher_JFA2018}
X.~Li and C.~Melcher.
\newblock Stability of axisymmetric chiral skyrmions.
\newblock {\em J. Funct. Anal.}, 275(10):2817--2844, 2018.

\bibitem{Braun_PRB1994}
H.-B. Braun.
\newblock Fluctuations and instabilities of ferromagnetic domain-wall pairs in
  an external magnetic field.
\newblock {\em Physical Review B}, 50(22):16485, 1994.

\bibitem{Zhou_NCOMM_2015}
Y.~Zhou, E.~Iacocca, A.~A. Awad, R.~K. Dumas, F.~C. Zhang, H.-B. Braun, and
  J.~{\AA}kerman.
\newblock Dynamically stabilized magnetic skyrmions.
\newblock {\em Nature communications}, 6:8193, 2015.

\bibitem{TejoRiverosChubykalo_srep2018}
F.~Tejo, A.~Riveros, J.~Escrig, K.~Y. Guslienko, and O.~Chubykalo-Fesenko.
\newblock Distinct magnetic field dependence of n\'eel skyrmion sizes in
  ultrathin nanodots.
\newblock {\em Sci. Rep.}, 8:6280, 2018.

\bibitem{TomaselloGuslienkoFinocchio_PRB2018}
R.~Tomasello, K.~Y. Guslienko, M.~Ricci, A.~Giordano, J.~Barker,
  M.~Carpentieri, O.~Chubykalo-Fesenko, and G.~Finocchio.
\newblock Origin of temperature and field dependence of magnetic skyrmion size
  in ultrathin nanodots.
\newblock {\em Phys. Rev. B}, 97:060402, Feb 2018.

\bibitem{KomineasMelcherVenakides_arXiv2019b}
Stavros Komineas, Christof Melcher, and Stephanos Venakides.
\newblock The profile of chiral skyrmions of large radius.
\newblock arXiv:1910.04818, 2019.

\bibitem{MantelMuratovSimon_PRB2019}
Anne Bernand-Mantel, Cyrill~B. Muratov, and Thilo~M. Simon.
\newblock Unraveling the role of dipolar versus dzyaloshinskii-moriya
  interactions in stabilizing compact magnetic skyrmions.
\newblock {\em Phys. Rev. B}, 101:045416, Jan 2020.

\bibitem{MantelMuratovSimon_arXiv2019}
Anne Bernand-Mantel, Cyrill~B. Muratov, and Thilo~M. Simon.
\newblock A quantitative description of skyrmions in ultrathin ferromagnetic
  films and rigidity of degree $\pm1$ harmonic maps from $\mathbb{R}^2$ to
  $\mathbb{S}^2$.
\newblock arXiv:1912.09854, 2019.

\bibitem{PapanicolaouTomaras_NPB1991}
N.~Papanicolaou and T.~N. Tomaras.
\newblock Dynamics of magnetic vortices.
\newblock {\em Nucl. Phys. B}, 360:425, 1991.

\bibitem{KomineasPapanicolaou_PhysD1996}
S.~Komineas and N.~Papanicolaou.
\newblock Topology and dynamics in ferromagnetic media.
\newblock {\em Physica D}, 99:81--107, 1996.

\bibitem{SchuetteGarst_PRB2014}
C.~Sch\"utte and M.~Garst.
\newblock Magnon-skyrmion scattering in chiral magnets.
\newblock {\em Phys. Rev. B}, 90:094423, Sep 2014.

\bibitem{BogdanovHubert_PSS1994}
A.~Bogdanov and A.~Hubert.
\newblock The properties of isolated magnetic vortices.
\newblock {\em Phys. Stat. Sol. B}, 186:527, 1994.

\bibitem{ChovanPapanicolaou_PRB2002}
J.~Chovan, N.~Papanicolaou, and S.~Komineas.
\newblock Intermediate phase in the spiral antiferromagnet
  {Ba}$_2${CuGe}$_2${O}$_7$.
\newblock {\em Phys. Rev. B}, 65:064433, Jan 2002.

\bibitem{BogdanovHubert_JMMM1999}
A.~N. Bogdanov and A.~Hubert.
\newblock The stability of vortex-like structures in uniaxial ferromagnets.
\newblock {\em J. Magn. Magn. Mater.}, 195:182, 1999.

\bibitem{BelavinPolyakov_JETP1975}
A.~A. Belavin and A.~M. Polyakov.
\newblock Metastable states of 2-dimensional isotropic ferromagnets.
\newblock {\em JETP Lett.}, 22:245, 1975.

\bibitem{KomineasPapanicolaou_PRB2015a}
S.~Komineas and N.~Papanicolaou.
\newblock Skyrmion dynamics in chiral ferromagnets.
\newblock {\em Phys. Rev. B}, 92:064412, Aug 2015.

\bibitem{AbramowitzStegun}
Milton Abramowitz and Irene~A. Stegun.
\newblock {\em Handbook of Mathematical Functions with Formulas, Graphs, and
  Mathematical Tables}.
\newblock Dover, New York, 1965.

\bibitem{Hoorfar_Hassani}
A.~Hoorfar and M.~Hassani.
\newblock Inequalities on the {L}ambert {$W$} function and hyperpower function.
\newblock {\em JIPAM. J. Inequal. Pure Appl. Math.}, 9(2):Article 51, 5, 2008.

\bibitem{Doering_Melcher_2017}
L.~D\"{o}ring and C.~Melcher.
\newblock Compactness results for static and dynamic chiral skyrmions near the
  conformal limit.
\newblock {\em Calc. Var. Partial Differential Equations}, 56(3):Art. 60, 30,
  2017.

\bibitem{Derrick}
G.~H. Derrick.
\newblock Comments on nonlinear wave equations as models for elementary
  particles.
\newblock {\em J. Mathematical Phys.}, 5:1252--1254, 1964.

\end{thebibliography}

\end{document}